\documentclass[final,1p,times]{elsarticle}

\usepackage[english]{babel}

\usepackage{amsmath}
\usepackage{amssymb}
\usepackage{float}
\usepackage{braket}
\usepackage{pdflscape}
\usepackage{siunitx}
\usepackage{resizegather}
\usepackage{mathtools}
\usepackage{graphicx,dblfloatfix}
\usepackage{enumitem}
\usepackage{subcaption}
\usepackage{lipsum}
\usepackage[colorlinks = true,
            linkcolor = blue,
            urlcolor  = blue,
            citecolor = blue,
            anchorcolor = blue]{hyperref}
\usepackage[font=small,labelfont=bf,
   justification=justified,
   format=plain]{caption}

\journal{JQSRT}
\begin{document}

\begin{frontmatter}

\title{High-resolution far-infrared spectroscopy and analysis of the $\nu_3$ and $\nu_6$ bands of chloromethane}

\author[1,2]{Pierre Hardy}
\author[1]{Cyril Richard}
\author[1]{Vincent Boudon}
\author[3]{Mohammad Vaseem Khan}
\author[4,5]{Laurent Manceron}
\author[6]{Nawel Dridi}

\affiliation[1]{organization={Laboratoire Interdisciplinaire Carnot de Bourgogne, UMR 6303 CNRS, Université de Bourgogne},
addressline={9 Av. A. Savary},
postcode={BP 47870},
city={F-21078 Dijon Cedex},
country={France}}

\affiliation[2]{organization={Institut UTINAM},
addressline={UMR 6213 CNRS–Univ. Franche-Comté},
postcode={BP 1615},
city={F-25010 Besançon Cedex},
country={France}}

\affiliation[3]{organization={CNRS, MONARIS, UMR 8233, Sorbonne Université},
addressline={4 Place Jussieu},
city={F-75005 Paris},
country={France}}

\affiliation[4]{organization={Synchrotron SOLEIL AILES Beamline},
postcode={B.P.48},
city={F-91192 Gif-sur-Yvette},
country={France}}

\affiliation[5]{organization={Université Paris Cité and Univ Paris Est Creteil, CNRS, LISA}, city={F-75013 Paris}, country={France}}

\affiliation[6]{organization={Université de Tunis, Laboratoire de Spectroscopie et Dynamique Moléculaire, Ecole Nationale Supérieure d’Ingénieurs de Tunis}, addressline={5 Av. Taha Hussein},city={1008 Tunis}, country={Tunisia}}
\begin{abstract}
Ro-vibrational spectra of the $\nu_3$ and $\nu_6$ bands of chloromethane ($\mathrm{CH_3Cl}$) were recorded in the 650--1130 $\mathrm{cm}^{-1}$ range using a Fourier transform spectrometer at the AILES beamline of the SOLEIL synchrotron facility. Two isotopologues ($\mathrm{CH_3^{35}Cl}$ and $\mathrm{CH_3^{37}Cl}$) have been analyzed with the tensorial formalism developed in Dijon and a total of 6753 lines were assigned. We derived 23 tensorial parameters for the lines positions (4 for the ground state, 6 for $\nu_3$, and 13 for $\nu_6$), and 7 for the lines intensities (4 for $\nu_3$, 3 for $\nu_6$). From those parameters and self-broadening coefficients found in the literature, we simulated spectra of both isotopologues. The derived parameters were converted in the Watson formalism to be compared with a previous study. Using these results, we set up a new database of calculated chloromethane spectral lines (ChMeCaSDa).
\end{abstract}
\begin{keyword}
Chloromethane \sep High-resolution infrared spectroscopy \sep Line positions \sep Tensorial formalism
\end{keyword}


\end{frontmatter}
\section{Introduction}
\begin{figure}[ht!]
    \centering
    \includegraphics[width=0.2\linewidth]{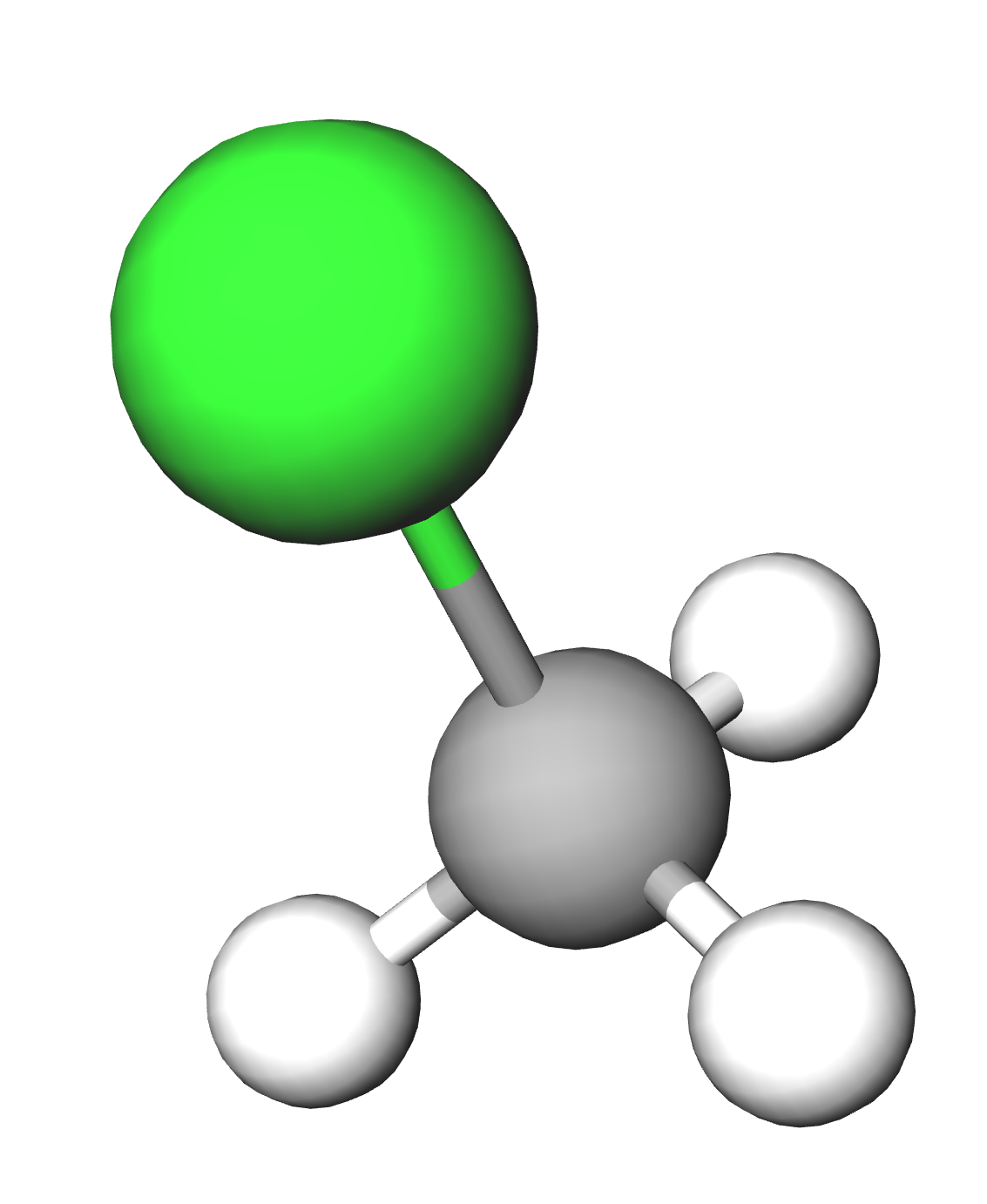}
    \caption{Three-dimensional representation of the Chloromethane molecule.}
    \label{fig:ch3cl3d}
\end{figure}

Chloromethane ($\mathrm{CH_3Cl}$, represented in Fig.~\ref{fig:ch3cl3d}) is a symmetric-top molecule belonging to the $C_{3v}$ symmetry group. It has six vibrational modes: three parallel vibrations of type $A_{1}$, and three degenerate pairs of perpendicular vibrations of type $E$.

This molecule is one of the simplest haloalkanes, belonging to the more general family of organohalogens which are abundant in the Earth's atmosphere, due to biological and industrial processes~\cite{oragnoearth}. Through catalytic processes, they lead to the destruction of atmospheric ozone~\cite{ozone}.

While $\mathrm{CH_3Cl}$ was detected for the first time in 1975 by Lovelock~\cite{lovelock}, its detection beyond the Earth's atmosphere is much more recent. Indeed, $\mathrm{CH_3Cl}$ was only detected in the surrounding of a protostar through submillimeter-wave spectroscopy in 2017, and in the coma of comet 67P/Churyumov-Gerasimenko through \textit{in situ} mass spectrometry~\cite{fayolle2017} the same year.

As one of the first molecules present at the early stages of solar systems it is of great astronomical interest, especially its primordial abundances which can be derived from cometary observations. As a matter of fact, comets are leftovers of the formation of our Solar system, which remained practically pristine since then.

\begin{figure}[t]
    \centering
    \includegraphics[width=0.65\linewidth]{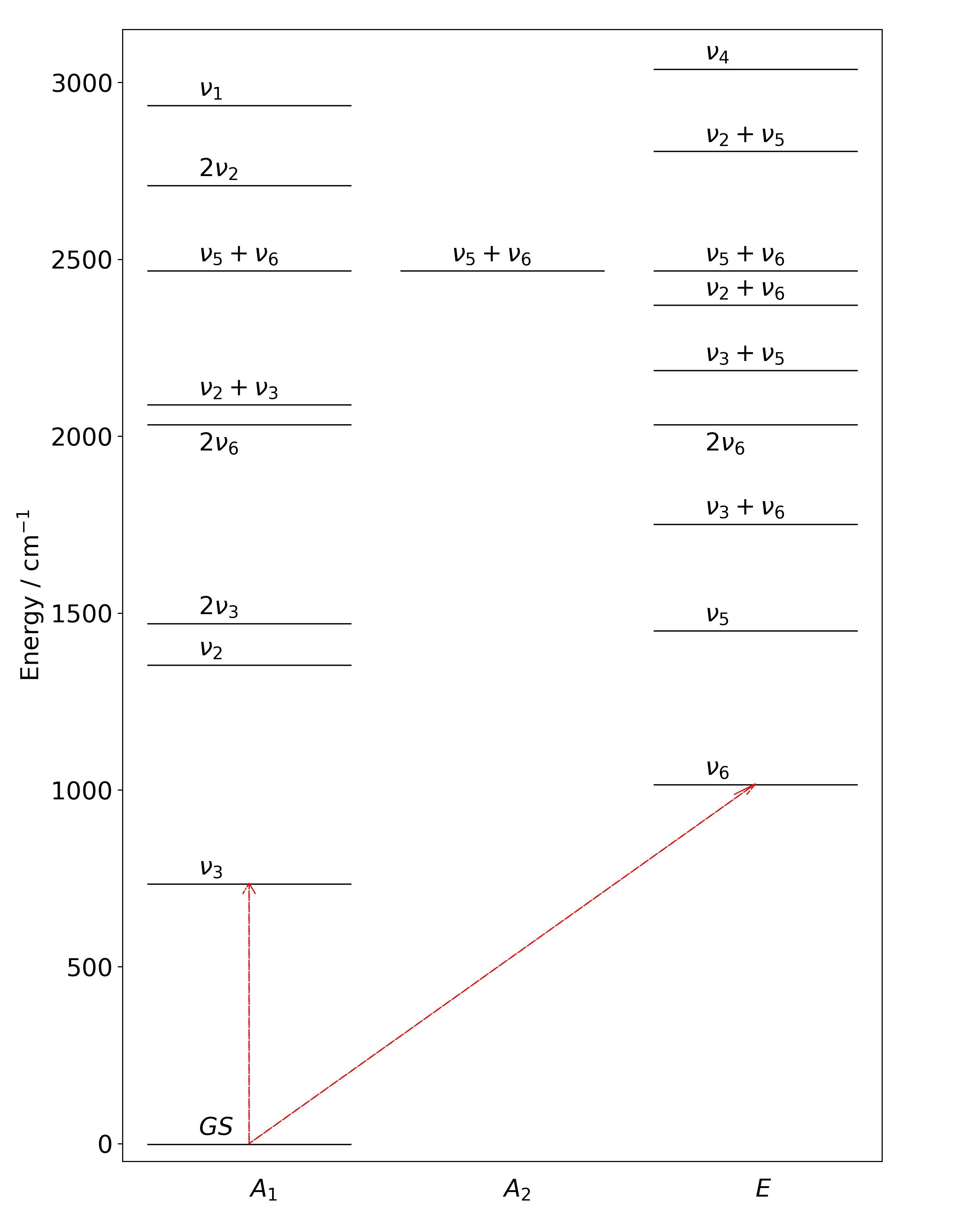}
    \caption{Energy level diagram of chloromethane. The vibrational states are separated in three columns, corresponding to their symmetry. The two transitions of interest are indicated by red arrows.}
    \label{fig:energy}
\end{figure}

Nonetheless, $\mathrm{CH_3Cl}$ is still to be observed in infrared wavelengths, and in this context, we analyzed the $\nu_3$ and $\nu_6$ bands of $\mathrm{CH_3^{35}Cl}$ and $\mathrm{CH_3^{37}Cl}$, located at 650--800 $\mathrm{cm}^{-1}$ ($\nu_3$) and 920--1150 $\mathrm{cm}^{-1}$ ($\nu_6$). Those two bands correspond to a C-Cl stretch and to a $\mathrm{CH_3}$ rock, respectively~\cite{shimanouchi}. Those two transitions are represented as red arrows in the energy level diagram of the molecule (Fig.~\ref{fig:energy}).

High-resolution spectra were recorded using the AILES beamline of the SOLEIL synchrotron facility.
The analyses and the modeling of spectra have been carried out using the Dijon tensorial formalism~\cite{boudon2011,wenger:hal-00362630} and the measurement of self-broadening coefficients of both bands, by considering previous works~\cite{DRIDI2019108,FATHALLAH2020106777}.
This paper is organized as follows. In section~\ref{sec:experimental}, we describe how the spectra were recorded at SOLEIL. Theoretical aspects of the tensorial formalism used are developed in section~\ref{sec:formalism}. In section~\ref{sec:discussion} we will discuss the results. Section \ref{sec:ChMeCaSDa} presents the new database built from our analyses (ChMeCaSDa), and we compare it with the HITRAN database of $\mathrm{CH_3Cl}$ in section \ref{sec:Comparison}. Finally, section~\ref{sec:ccl} presents our conclusions.

\section{Experimental details}\label{sec:experimental}
The experiment that took place at the AILES beamline of the SOLEIL synchrotron facility was already described in~\cite{FATHALLAH2020106777}.
The spectra of pure $\mathrm{CH_{3}Cl}$ were recorded from 650 to 1130 $\mathrm{cm}^{-1}$ at room temperature (T = 295~K) and at seven pressures ranging from 1.02 to 10.24~mbar, using a Bruker IFS 125 HR Fourier transform spectrometer. These are the same as those described previously in ref~\cite{ROY2006139}. The spectrometer was equipped with a Globar source, a Ge/KBr beamsplitter, and a 4~K-cooled HgCdTe photoconductivity detector which provided a signal-to-noise ratio eight times larger than commercial detectors in this range~\cite{faye}. The average signal-to-noise ratio for the seven spectra vary between 700-1500 for $\nu_3$~\cite{DRIDI2020107036}, and 300-600 for $\nu_6$~\cite{FATHALLAH2020106777}.
\color{black}Its aperture diameter was 1.7~mm.
Methyl chloride in natural abundance was provided by Air Liquide with a stated purity of 99.5~\% used.
Since no purity accuracy was specified by the manufacturer, no correction was made.
The $\mathrm{CH_{3}Cl}$ gas was contained in an absorption cell with a path length of $24.4 \pm 1$~cm equipped with a ZnSe window with anti-reflective coating.
The temperature of the gas was measured using platinum probes with an estimated accuracy of $\pm1$~K, and the pressure was measured with a Pfeiffer temperature-controlled gauge with a reading accuracy of 0.2~\%, reaching 0.5~\% by taking into account the uncertainty arising from small fluctuations in the pressure during the recording.
To minimize the effects of the atmospheric gases in the beam path, the spectrometer was continuously evacuated below a pressure of 2.3 $\times 10^{-4}$~mbar by a dry pump.

For each pressure, an empty cell spectrum, recorded under the same conditions and orientation, was used as a reference to limit baseline variations that could introduce errors in the measured parameters.
The resolution was set to 0.003~$\mathrm{cm}^{-1}$, which corresponds to a maximum optical path difference of 300~cm.
The interferograms were Fourier transformed using the method described in the OPUS package of the Bruker software \footnote{\url{https://www.bruker.com/en/products-and-solutions/infrared-and-raman/gas-analysis/opus-ga-gas-analysis-software.html}} without apodization, using a Mertz-phase error correction method~\cite{mertz} and a zero-filling factor of 4.
A summary of the experimental conditions is presented in Table \ref{tab:expe}.\color{black}

\begin{table}[h!]
\caption{Experimental conditions under which the seven spectra of $\mathrm{CH_{3}Cl}$ were recorded at the AILES beamline of the SOLEIL synchrotron facility. The Signal/Noise values are RMS values around 785 $\mathrm{{cm}^{-1}}$.}
\resizebox{\textwidth}{!}{%
\begin{tabular}{rrrr}
\hline
Spectrum number      & $\mathrm{CH_{3}Cl}$ pressure (mbar)  & Signal/Noise (RMS) & Number of Scans   \\ \hline
1 & 1.020 $\pm$ 0.005 & 1415 & 750\\
2 & 2.100 $\pm$ 0.011 & 1198 & 150\\
3 & 3.550 $\pm$ 0.018 & 929 & 960\\
4 & 5.050 $\pm$ 0.025 & 1176 & 780\\
5 & 6.580 $\pm$ 0.033 & 1366 & 840\\
6 & 8.520 $\pm$ 0.043 & 1219 & 900\\
7 & 10.240 $\pm$ 0.051 & 1177 & 940\\ \hline
\end{tabular}%
}
\label{tab:expe}
\end{table}

\section{Theoretical aspects}\label{sec:formalism}
\subsection{Tensorial formalism}
The molecule of chloromethane being highly symmetric, the tensorial formalism developed in Dijon~\cite{boudon2011,wenger:hal-00362630} was used to derive spectroscopic parameters. Originally developed for highly symmetric (spherical tops) species, this formalism has since been extended to other symmetries, such as the symmetric-top $C_{3v}$~\cite{NIKITIN1997,EBL05a,EBL05b}. For example, some infrared bands of trioxane~\cite{richard:trioxane} and germane~\cite{germane} were recently studied using this formalism. We will briefly review the main ideas of this method. A more complete description is available in Ref.~\cite{EBL05b}.

The vibrational levels of $\mathrm{CH_{3}Cl}$ are grouped in several polyads $P_k$, and the Hamiltonian is expressed as a sum of operators specific to each polyad:
\begin{equation}
    \mathcal{H}=\mathcal{H}_{\{P_0=GS\}}+\mathcal{H}_{\{P_1\}}+\mathcal{H}_{\{P_2\}}+ \dots
\end{equation}

The effective Hamiltonian for a given vibrational polyad is defined as
\begin{equation}
    {\tilde{\mathcal{H}}}^{<polyad>}=P^{<polyad>}\mathcal{H}P^{<polyad>},
\end{equation}
where
\begin{equation}
    P^{<polyad>}=\sum_{i}\ket{\psi^{i}_{v}}\bra{\psi^{i}_{v}}
\end{equation}
is the projection operator on the vibrational Hilbert subspace, and index $i$ runs over the vibrational states of the polyad. For a given polyad $P_n$, the effective Hamiltonian is written as
\begin{equation}
    {\tilde{\mathcal{H}}}^{<P_n>}={\tilde{\mathcal{H}}}^{<P_n>}_{\{GS\}}+{\tilde{\mathcal{H}}}^{<P_n>}_{\{P_1\}}+\dots+{\tilde{\mathcal{H}}}^{<P_n>}_{\{P_n\}}.
\end{equation}
In this work we will use the following effective Hamiltonians:
\begin{equation}
    {\tilde{\mathcal{H}}}^{<GS>}={\tilde{\mathcal{H}}}^{<GS>}_{\{GS\}}
\end{equation}
and
\begin{equation}
    {\tilde{\mathcal{H}}}^{<\nu_j>}={\tilde{\mathcal{H}}}^{<\nu_j>}_{\{GS\}}+{\tilde{\mathcal{H}}}^{<\nu_j>}_{\{\nu_j\}},
\end{equation}
with $j=3,6$. These effective Hamiltonians are then expanded a sum of coupled rotational and vibrational operators themselves built on elementary creation, annihilation and angular momentum operators. The detailed tensorial construction, based on an extensive use of symmetry properties and group theoretical method has already been described many times and can be found, for instance in Refs.~\cite{EBL05b,hrs021}. In the same papers, we also explained the construction of the effective dipole moment operator using identical methods, for the calculation of line intensities.

\subsection{Watson formalism}
In the Watson formalism, the Hamiltonian for $C_{3v}$ symmetry molecules can be written as: 
\begin{equation}
\begin{array}{ll}
    \tilde{\mathcal{H}}_{Watson}= & T_v +BJ(J+1) +(C-B)K^2 \\
    & -D_J J^2(J+1)^2 -D_{JK}J(J+1)K^2 -D_K K^4 \\
   & +H_J J^3(J+1)^3 +H_{JK}J^2(J+1)^2K^2+ H_{KJ}J(J+1)K^4+H_K K^6 \\
  &  -2C\zeta Kl+\eta_J J(J+1)Kl+\eta_K K^3 l\\
  &+\frac{1}{2}(q_+ +D_{q_J} J(J+1)+D_{q_K} K^2)({L_+}^2 {J_-}^2 + {L_-}^2 {J_+}^2).
\end{array}
\end{equation}
Here, $T_v$ is the vibrational term. $B$, $C$, $D_J$, $D_{JK}$, $D_K$, $H_J$, $H_{JK}$, $H_{KJ}$ and $H_K$ are the traditional rotational and centrifugal distortion constants, while $\zeta$, $\eta_J$ and $\eta_K$ are the Coriolis interaction constants for doubly degenerate vibrations that, in the present case, only exist for the $\nu_6$ mode. Finally, $q$, $D_{qJ}$ and $D_{qK}$ are $l$-doubling constants. We purposely omitted here unused terms, like pure rotational $C_{3v}$ splitting terms $\varepsilon$, $\varepsilon_J$ ,\ldots , that were not determined in the present study.

Going from one formalism to the other is possible by expanding the two different Hamiltonians in terms of elementary rotational operators, as explained in \cite{willis} and as it was applied in~\cite{richard:trioxane} for another molecule with $C_{3v}$ symmetry.

\section{Analysis and discussion}\label{sec:discussion}

\subsection{Lines positions}

As a starting point in the fitting procedure, we used the ground state rotational data of $\mathrm{CH_3^{35}Cl}$ from a previous study~\cite{DEMAISON1994147}. From those parameters, an initial synthetic spectrum was obtained. Then we assigned lines using SPVIEW~\cite{wenger:hal-00362630}  and the first experimental spectrum (with a pressure of 1.020 mbar, see Table \ref{tab:expe}). This spectrum was chosen among the others to avoid saturated lines that might occur when dealing with higher pressures. \color{black} A standard iterative Levenberg-Marquardt non-linear least squares fit was finally performed, with the XTDS software~\cite{ELHILALI20101305} using the $C_{3v}$ package for spectra modeling and job executions. For each isotopologue, we performed a global fit: lines positions in ground state and in the $\nu_3$ and $\nu_6$ bands were fitted simultaneously.

In the ground state, 558 lines have been fitted for $\mathrm{CH_3^{35}Cl}$. 143 of those lines are in microwave frequencies, while 415 are in the far-infrared region. The root mean squares deviations obtained for the fit were respectively of 0.358~MHz in the microwave and $0.162 \times 10^{-3}\mathrm{cm^{-1}}$ in the infrared.

\begin{figure*}[p]
    \centering
    \includegraphics[width=0.8\linewidth]{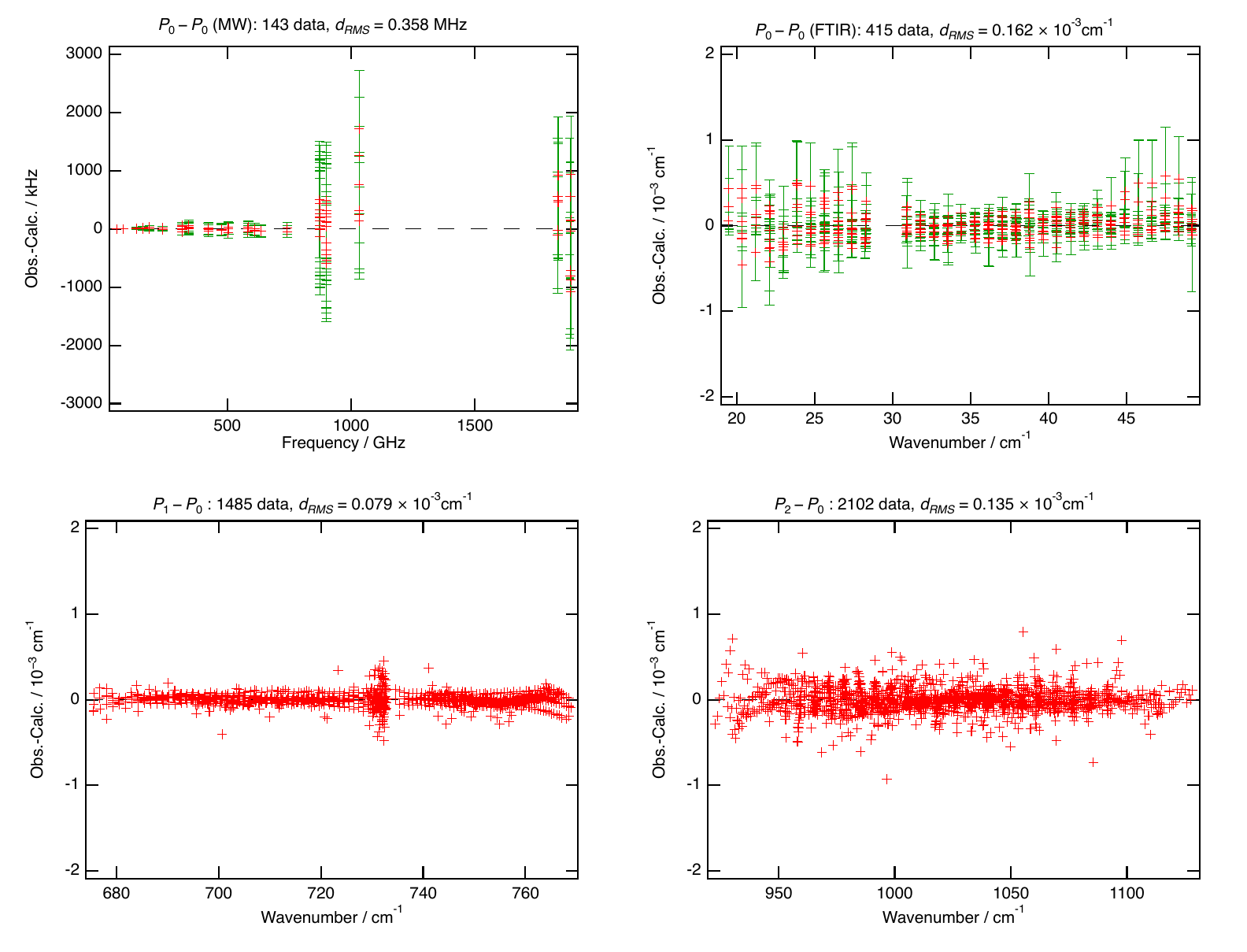}
    \captionof{figure}{Fit residuals (in red) for the line positions of $\mathrm{CH_3^{35}Cl}$ for the ground state, the $\nu_3$ band, and the $\nu_6$ band. Two sets of data were used for the ground state: microwave (upper left panel) and far-infrared (upper right panel). The experimental errors are represented by vertical green lines in the two first panels (ground state). For the two excited bands, the experimental errors are constant and equal to $10^{-3}\ \mathrm{cm}^{-1}$ and are not displayed.}
    \label{fig:errors35}
\end{figure*}

For the $\nu_3$ and $\nu_6$ bands of $\mathrm{CH_3^{35}Cl}$, we have respectively fitted 1405 and 2102 lines, and the derived root mean squares deviations were 0.079 $\times 10^{-3}\mathrm{cm}^{-1}$ and 0.135 $\times 10^{-3}\mathrm{cm}^{-1}$. The fit residuals of the two excited bands are represented on the bottom part of Figure~\ref{fig:errors35}. For better visibility, the error bars were not represented on the plots since they are equal to $10^{-3}\mathrm{cm^{-1}}$ for each fitted line. 
In the same way, 127 lines of $\mathrm{CH_3^{37}Cl}$ were used in the ground state (only microwave), 1153 in the $\nu_3$ band, and 1328 in the $\nu_6$ band in our fit. The obtained root mean squares deviations are respectively 357.207~kHz, 0.114 $\times 10^{-3}\mathrm{cm}^{-1}$, and 0.182 $\times 10^{-3}\mathrm{cm}^{-1}$.

\begin{figure*}[p]
    \centering
    \includegraphics[width=0.8\linewidth]{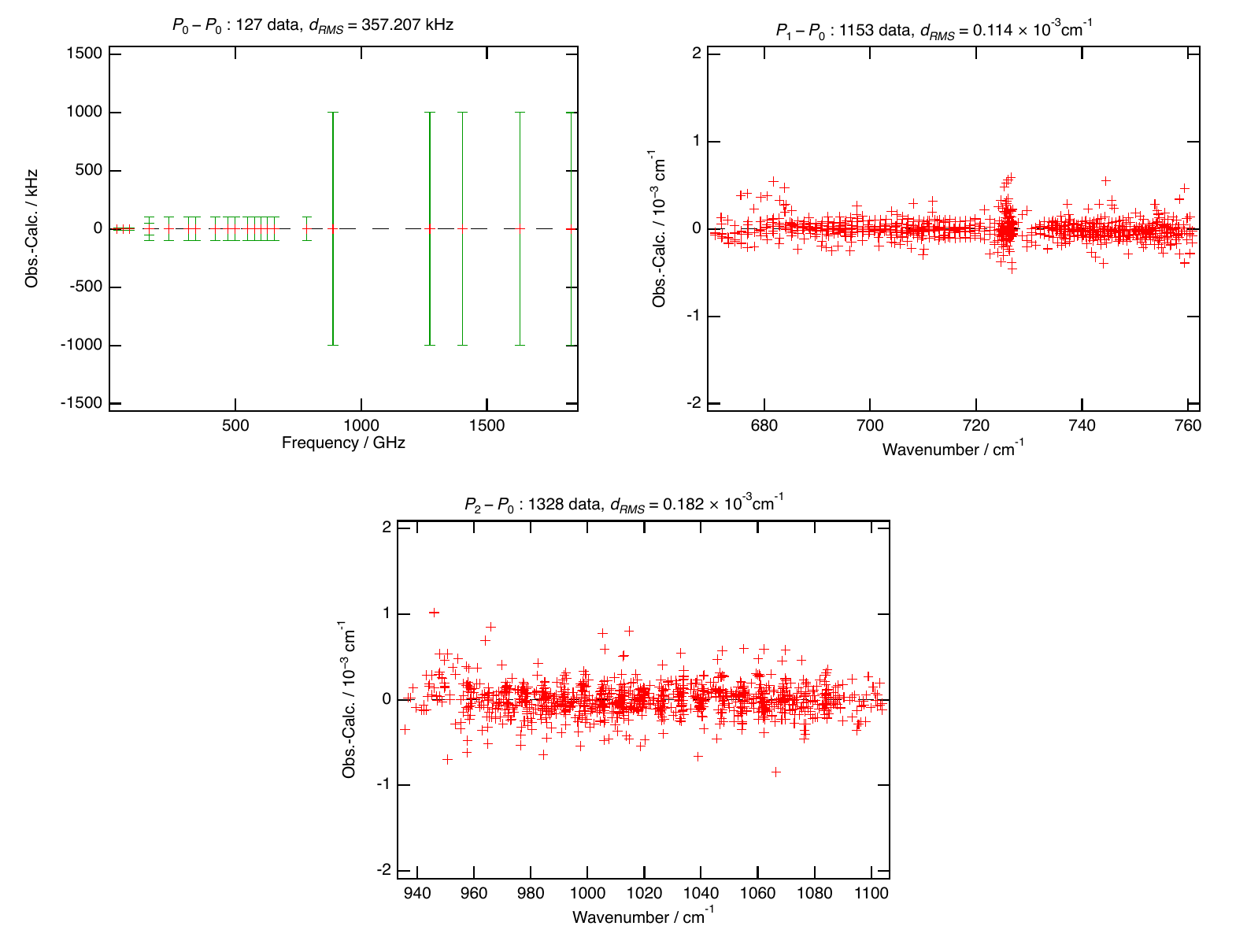 }
    \caption{Fit residuals (in red) of the line positions of $\mathrm{CH_3^{37}Cl}$ for the ground state, the $\nu_3$ band, and the $\nu_6$ band. The experimental errors are represented by vertical green lines in the first panel (ground state). For the two excited bands, the experimental errors are constant and equal to $10^{-3}\ \mathrm{cm}^{-1}$ and are not displayed.}
    \label{fig:errors37}
\end{figure*}

In total, 23 free parameters (4 for the ground state, 6 for $\nu_3$, and 13 for $\nu_6$) have been fitted for the line position of both isotopologues.
The results and fit statistics for both isotopologues are presented in Tables~\ref{tab:results35} and \ref{tab:results37}.
Figures~\ref{fig:errors35} and \ref{fig:errors37} present the fit residuals for line positions of $\mathrm{CH_{3}^{35}Cl}$ and $\mathrm{CH_{3}^{37}Cl}$, respectively. $P_0$, $P_1$, and $P_2$ correspond to the different polyads of the molecule (corresponding to the ground state, the $\nu_3=1$, and $\nu_6=1$ excited state).

\subsection{Lines intensities}\label{sec:int}
Similarly to a previous work dedicated to analyzing two bands of germane~\cite{germane}, we performed a fit of the dipole moment operators for both isotopologues up to the second order. SPVIEW was used again to assign and select lines for a non-linear least squares fit.  Contrary to the fit of lines positions, the seven spectra were used for the fit of the dipole moment operators. The experimental lines intensities were retrieved from previous works (\cite{DRIDI2020107036} for $\nu_3$;~\cite{FATHALLAH2020106777} for $\nu_6$). \color{black}For the $\nu_3$ band, we fitted 827 lines of $\mathrm{CH_3^{35}Cl}$ and 481 lines of $\mathrm{CH_3^{37}Cl}$. The root mean squares deviations of our fit are 12.185~\% for $\mathrm{CH_3^{35}Cl}$ and 14.691~\% for $\mathrm{CH_3^{37}Cl}$.
For the $\nu_6$ band, we fitted 1849 lines of $\mathrm{CH_3^{35}Cl}$ and 741 lines of $\mathrm{CH_3^{37}Cl}$, with root mean squares deviations equal to 6.612~\% and 7.695~\%, respectively. We fitted 7 tensorial parameters  (4 for $\nu_3$, 3 for $\nu_6$) for both isotopologues. The fit parameters are summarized in Table~\ref{tab:int}, where $\Omega$ is the
maximum degree of the rotational operator, K its rank, C an irreducible representation of $C_{3v}$, and $n$ a
multiplicity index. More details about the notation is presented in the review of Boudon \textit{et al.} \cite{BOUDON2004620}. \color{black}

\subsection{Self-broadening coefficients}
\begin{figure}[h!]
    \centering
    \includegraphics[width=0.49\linewidth]{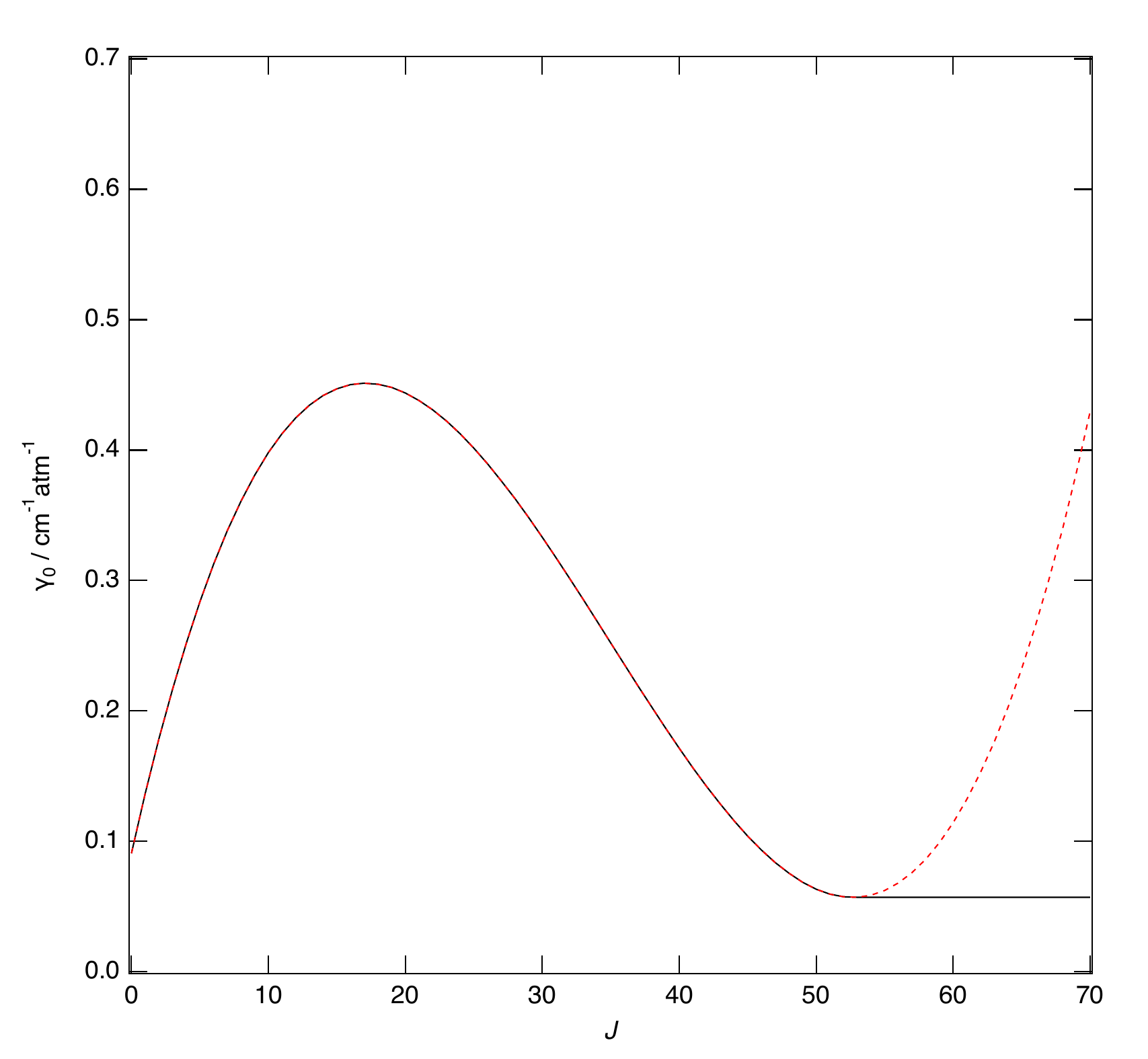}
    \caption{Self-broadening coefficients $\gamma_0$ of the $\nu_3$ band as a function of the quantum number $J$, with $K$ set to 12. The red-dashed curve represents the polynomial fit as expressed in~\cite{DRIDI2019108}. The black curve represents the function $\gamma_0(J)$ used in this work, to avoid an infinite divergence of the coefficients as $J$ increases. $\gamma_0$ is expressed in units of $\mathrm{cm}^{-1}\mathrm{atm}^{-1}$.}
    \label{fig:broad1}
\end{figure}

\begin{figure}[h!]
    \centering
    \includegraphics[width=0.49\linewidth]{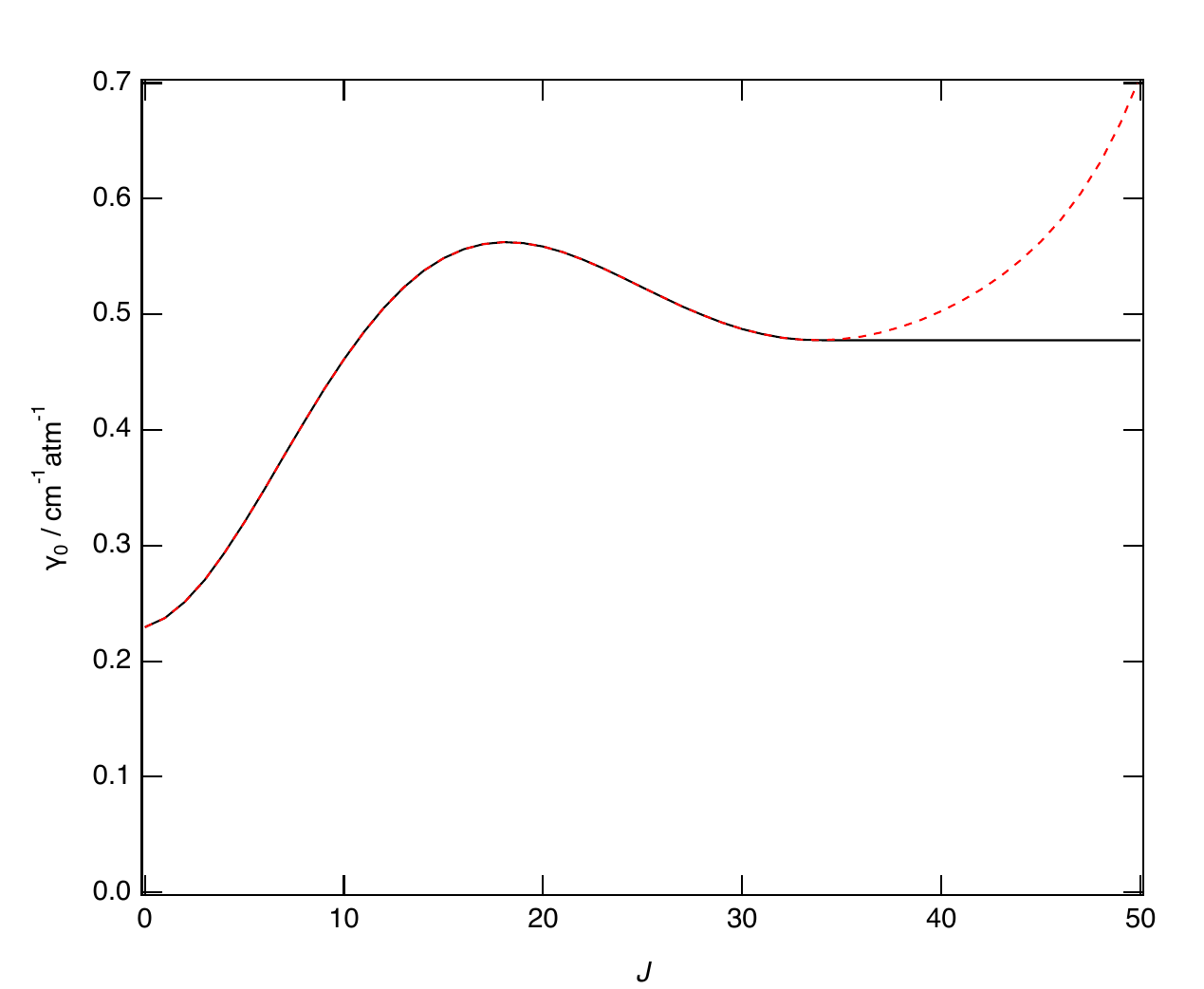}
    \caption{Self-broadening coefficients $\gamma_0$ of the $\nu_6$ band as a function of the quantum number $J$, with K set to 4. The red-dashed curve represents the polynomial fit as expressed in~\cite{FATHALLAH2020106777}. The black curve represents the function $\gamma_0(J)$ used in this work, to avoid an infinite divergence of the coefficients as $J$ increases. $\gamma_0$ is expressed in units of $\mathrm{cm}^{-1}\mathrm{atm}^{-1}$.}
    \label{fig:broad2}
\end{figure}

The self-broadening coefficients of chloromethane were recently studied by Dridi {\em et al.\/}~\cite{DRIDI2019108} and Fatallah {\em et al.\/}~\cite{FATHALLAH2020106777}, respectively, for the $\nu_3$ band and $\nu_6$ band. In each case, polynomial fits of the broadening coefficients were derived. They are expressed as:
\begin{equation*}
    \gamma_0(J,K)=\left\{
    \begin{array}{ll}
    A_0+A_1 J+A_2 J^2 +A_3 K^2 + A_4 J^3 & \mbox{($\nu_3$)}\\
    \\
    A_j^{00}+A_j^{01}J+A_j^{02}J^2+A_j^{03}J^3\\[3pt]
    +A_j^{04}J^4+A_j^{05}J^5+A_j^{06}J^6+(A_j^{20}+A_j^{21}e^{-A_j^{22}(J-J_0)})K^2& \mbox{($\nu_6$)}
    
    \end{array}
\right.
\end{equation*}
with A-parameters expressed in $\mathrm{cm^{-1}atm^{-1}}$ and $J_0$ unitless. The values of these parameters are summarized in Table~\ref{tab:parabroad}. Those polynomials were obtained with $2\leq K\leq 13$ and $2\leq J\leq 59$ for $\nu_3$; $0\leq K\leq 12$ and $1\leq J\leq 55$ for $\nu_6$.
Assigning a value of $K$ for each line of the spectrum is not straightforward in the Dijon formalism (since it does not appear explicitly, see for instance Eq. 12 in Ref. \cite{richard:trioxane}), thus we fixed the value of $K$ to a constant value for each band ($K=12$ for the $\nu_3$ band, $K=4$ for the $\nu_6$ band). 
Moreover, to avoid the divergence of extrapolated $\gamma_0$ at large $J$ values, we decided to keep these self-broadening coefficients constant once the minimum is reached (see Figures~\ref{fig:broad1} and \ref{fig:broad2}).

\begin{table*}[p]
\caption{Effective Hamiltonian parameters for the ground-vibrational state, $\nu_3$, and $\nu_6$ bands of $\mathrm{CH_3^{35}Cl}$. Values in parentheses represent the standard deviation, in units of the last two digits. No standard deviation indicates that this parameter was fixed.}
\centering
\resizebox{0.8\textwidth}{!}{\begin{tabular}{lllllll}
\hline
\hline
Parameters            & \multicolumn{6}{l}{Value/$\mathrm{cm^{-1}}$ (Hamiltonian $\Tilde{H}$)}                                      \\
$t_{i}^{\Omega(K,nC)}$ & \multicolumn{2}{l}{GS} & \multicolumn{2}{l}{$\nu_{3}$} & \multicolumn{2}{l}{$\nu_{6}$}         \\ \hline
$t_{i}^{0(0,S^{+})}$ &                   &                   & 732.842105(74) &                  & 1018.071457(63) &                  \\
$t_{i}^{1(1,S^{-})}$ &                   &                   &                &                  &   -1.8535547(99)&                  \\
$t_{i}^{2(0,S^{+})}$ &  2.0308384259(22) &                   &  -5.43068(83)  & $\times 10^{-3}$ &    7.53316(81)  & $\times 10^{-3}$ \\
$t_{i}^{2(2,S^{+})}$ &  9.7210192000     & $\times 10^{-1}$  &  -9.7182(49)   & $\times 10^{-4}$ &    7.93968(67)  & $\times 10^{-3}$ \\
$t_{i}^{2(2,D)}$     &                   &                   &                &                  &    2.4462(38)   & $\times 10^{-4}$ \\
$t_{i}^{3(1,S^{-})}$ &                   &                   &                &                  &    3.5398(29)   & $\times 10^{-5}$ \\
$t_{i}^{3(3,S^{-})}$ &                   &                   &                &                  &   -2.4482(16)   & $\times 10^{-5}$ \\
$t_{i}^{4(0,S^{+})}$ & -2.814049(14)     & $\times 10^{-6}$  &   1.292(33)    & $\times 10^{-7}$ &   -5.991(36)    & $\times 10^{-7}$ \\
$t_{i}^{4(2,S^{+})}$ &  5.8614235034     & $\times 10^{-7}$  &  -4.98(12)     & $\times 10^{-8}$ &    3.122(19)    & $\times 10^{-7}$ \\
$t_{i}^{4(2,D)}$     &                   &                   &                &                  &    3.59(23)     & $\times 10^{-7}$ \\
$t_{i}^{4(4,S^{+})}$ &                   &                   &   1.994(48)    & $\times 10^{-8}$ &   -1.1572(71)   & $\times 10^{-7}$ \\
$t_{i}^{4(4,D)}$     &                   &                   &                &                  &   -5.47(34)     & $\times 10^{-7}$ \\
$t_{i}^{5(1,S^{-})}$ &                   &                   &                &                  &    8.5(3.2)     & $\times 10^{-12}$\\
$t_{i}^{6(0,S^{+})}$ &  6.5856(15)       & $\times 10^{-11}$ &                &                  &                 &                  \\
$t_{i}^{6(2,S^{+})}$ &  1.0677848338     & $\times 10^{-11}$ &                &                  &                 &                  \\
$t_{i}^{6(4,S^{+})}$ & -4.0459(23)       & $\times 10^{-12}$ &                &                  &                 &                  \\ \hline
Lines fitted         & 558               &                   & 1485           &                  & 2102            &                  \\
$J_{max}$            & 71                &                   & 53             &                  & 45              &                  \\
Free parameters      & 4                 &                   & 6              &                  & 13              &                  \\
$d_{\text{RMS}}$     & 0.358 MHz (MW)    &                   & 0.079 $\times 10^{-3}\mathrm{cm}^{-1}$       &                  & 0.135 $\times 10^{-3}\mathrm{cm}^{-1}$        &                  \\
                     & 0.162 $\times 10^{-3}\mathrm{cm}^{-1}$ (FTIR) & &      &                  &                 &                  \\ \hline
Total of lines       & \multicolumn{6}{l}{4145}\\
Standard deviation   & \multicolumn{6}{l}{0.303}\\ \hline \hline
\end{tabular}}

\label{tab:results35}
\end{table*}

\clearpage

\begin{table*}[p]
\caption{Effective Hamiltonian parameters for the ground-vibrational state, $\nu_3$, and $\nu_6$ bands of $\mathrm{CH_3^{37}Cl}$. Values in parentheses represent the standard deviation, in units of the last two digits. No standard deviation indicates that this parameter was fixed.}
\centering
\resizebox{0.8\textwidth}{!}{\begin{tabular}{lllllll}
\hline
\hline
Parameters            & \multicolumn{6}{l}{Value/$\mathrm{cm^{-1}}$ (Hamiltonian $\Tilde{H}$)}                                      \\
$t_{i}^{\Omega(K,nC)}$ & \multicolumn{2}{l}{GS} & \multicolumn{2}{l}{$\nu_{3}$} & \multicolumn{2}{l}{$\nu_{6}$}         \\ \hline
$t_{i}^{0(0,S^{+})}$ &                   &                   & 727.029509(82) &                  & 1017.682771(80) &                  \\
$t_{i}^{1(1,S^{-})}$ &                   &                   &                &                  &   -1.848751(15) &                  \\
$t_{i}^{2(0,S^{+})}$ &  2.026286226(18)  &                   &  -5.3440(11)   & $\times 10^{-3}$ &    7.5379(14)   & $\times 10^{-3}$ \\
$t_{i}^{2(2,S^{+})}$ &  9.7349573918     & $\times 10^{-1}$  &  -9.7755(63)   & $\times 10^{-4}$ &    7.9208(12)   & $\times 10^{-3}$ \\
$t_{i}^{2(2,D)}$     &                   &                   &                &                  &    2.4057(61)   & $\times 10^{-4}$ \\
$t_{i}^{3(1,S^{-})}$ &                   &                   &                &                  &    3.5568(65)   & $\times 10^{-5}$ \\
$t_{i}^{3(3,S^{-})}$ &                   &                   &                &                  &   -2.4521(39)   & $\times 10^{-5}$ \\
$t_{i}^{4(0,S^{+})}$ & -2.737719(21)     & $\times 10^{-6}$  &   1.300(45)    & $\times 10^{-7}$ &   -6.050(98)    & $\times 10^{-7}$ \\
$t_{i}^{4(2,S^{+})}$ &  5.7058387143     & $\times 10^{-7}$  &  -5.00(17)     & $\times 10^{-8}$ &    3.158(52)    & $\times 10^{-7}$ \\
$t_{i}^{4(2,D)}$     &                   &                   &                &                  &    3.67(46)     & $\times 10^{-7}$ \\
$t_{i}^{4(4,S^{+})}$ &                   &                   &   1.994(63)    & $\times 10^{-8}$ &   -1.175(20)    & $\times 10^{-7}$ \\
$t_{i}^{4(4,D)}$     &                   &                   &                &                  &   -5.60(71)     & $\times 10^{-7}$ \\
$t_{i}^{5(1,S^{-})}$ &                   &                   &                &                  &    1.31(50)     & $\times 10^{-11}$\\
$t_{i}^{6(0,S^{+})}$ &  6.51389(37)      & $\times 10^{-11}$ &                &                  &                 &                  \\
$t_{i}^{6(2,S^{+})}$ &  1.0564481018     & $\times 10^{-11}$ &                &                  &                 &                  \\
$t_{i}^{6(4,S^{+})}$ & -4.00696(43)      & $\times 10^{-12}$ &                &                  &                 &                  \\ \hline
Lines fitted         & 127               &                   & 1153           &                  & 1328            &                  \\
$J_{max}$            & 71                &                   & 53             &                  & 43              &                  \\
Free parameters      & 4                 &                   & 6              &                  & 13              &                  \\
$d_{\text{RMS}}$     & 0.369 MHz (MW)    &                   & 0.114 $\times 10^{-3}\mathrm{cm}^{-1}$       &                  & 0.167 $\times 10^{-3}\mathrm{cm}^{-1}$        &                  \\ \hline 
Total of lines       & \multicolumn{6}{l}{2608}\\
Standard deviation   & \multicolumn{6}{l}{0.193}\\ \hline \hline

\end{tabular}
}

\label{tab:results37}
\end{table*}

\clearpage

\begin{table*}[t]
\caption{Effective dipole moment parameter values for the $\nu_3$ and $\nu_6$ bands of chloromethane, for both isotopologues. Standard deviation is indicated in parenthesis, in the unit of the last two digits.  $\Omega$ is the maximum degree of the rotational operator, K its rank, C an irreducible representation of $C_{3v}$, and $n$ is a multiplicity index. More details about the tensorial formalism and the notation used here can be found in the review of Boudon \textit{et al.} \cite{BOUDON2004620}.}
\centering
\small\begin{tabular}{lllllrr}
\hline \hline
Band & Order & $\Omega$ (K, nC) & \{s\} $C_1$& \{$s^{\prime}$\} $C_2$& \multicolumn{2}{c}{Value / Debye} \\ 
     &       &                  &           &           & $\mathrm{CH_3^{35}Cl}$      & $\mathrm{CH_3^{37}Cl}$                                    \\ \hline
$\nu_3$   & 0     & 0(0,S+)          & 000000 S+ & 001000 S+ & 3.6682(41) $\times 10^{-1}$ & 3.6372(18) $\times 10^{-1}$                              \\
$\nu_3$   & 1     & 1(1,P)           & 000000 S+ & 001000 S+ & 1.971(60) $\times 10^{-4}$  & 2.340(28) $\times 10^{-4}$                               \\
$\nu_3$   & 2     & 2(0,S+)          & 000000 S+ & 000000 S+ & -2.05(12) $\times 10^{-6}$  & 6.77(64) $\times 10^{-7}$                              \\
$\nu_3$   & 2     & 2(2,P)           & 000000 S+ & 001000 S+ & 4.65(69) $\times 10^{-6}$   & 6.6(6.0) $\times 10^{-7}$                              \\ \hline
\multicolumn{2}{l}{Number of lines} &  &  &  & 827 & 481 \\
\multicolumn{2}{l}{Root mean square} &  &  &  & 12.185\% & 14.691\% \\ \hline
$\nu_6$   & 0     & 0(0,S+)          & 000000 S+ & 000001 P  & 1.82289(72) $\times 10^{-1}$ & 5.8156(28) $\times 10^{-2}$                             \\
$\nu_6$   & 1     & 1(1,S-)          & 000000 S+ & 000001 P  & -7.971(69) $\times 10^{-4}$  & -1.903(33) $\times 10^{-4}$                              \\
$\nu_6$   & 1     & 1(1,P)           & 000000 S+ & 000001 P  & 1.418(21) $\times 10^{-4}$  & 3.09(11) $\times 10^{-5}$                              \\ \hline
\multicolumn{2}{l}{Number of lines} &  &  &  & 1670 & 694 \\
\multicolumn{2}{l}{Root mean square} &  &  &  & 6.612\% & 7.695\% \\
\hline\hline               
\end{tabular}
\label{tab:int}
\end{table*}

\begin{table}[h!]
\caption{Parameters used for the self-broadening coefficients of the two bands. All values (except $J_0$) are expressed in $\mathrm{cm^{-1}atm^{-1}}$. Retrieved from Dridi {\em et al.\/}~\cite{DRIDI2019108} and Fatallah {\em et al.\/}~\cite{FATHALLAH2020106777}.}
\begin{center}
\begin{tabular}{llll}
\hline \hline
$A_0$ & 0.16545 & $A_J^{03}$ & -0.00090 \\[1pt]
$A_1$ & 0.0473 & $A_J^{04}$ & $2.07872 \times 10^{-5}$ \\[1pt]
$A_2$ & $-1.833 \times 10^{-3}$ & $A_J^{05}$ & $-4.11387 \times 10^{-7}$  \\[1pt]
$A_3$ & $-0.511 \times 10^{-3}$ & $A_J^{06}$ & $2.38023 \times 10^{-9}$ \\[1pt]
$A_4$ & $1.752 \times 10^{-5}$ & $A_J^{20}$ & -0.00023  \\[1pt]
$A_J^{00}$ & 0.53272  & $A_J^{21}$ & -0.00846  \\[1pt]
$A_J^{01}$ & -0.0745 & $A_J^{22}$ & 0.26497 \\[1pt]
$A_J^{02}$ & 0.01359& $J_0$ & 3\\ \hline \hline
\end{tabular}\end{center}

\label{tab:parabroad}
\end{table}

\newpage
\subsection{Simulation of spectra}
\begin{figure*}[h!]
    \centering
    \includegraphics[width=\linewidth]{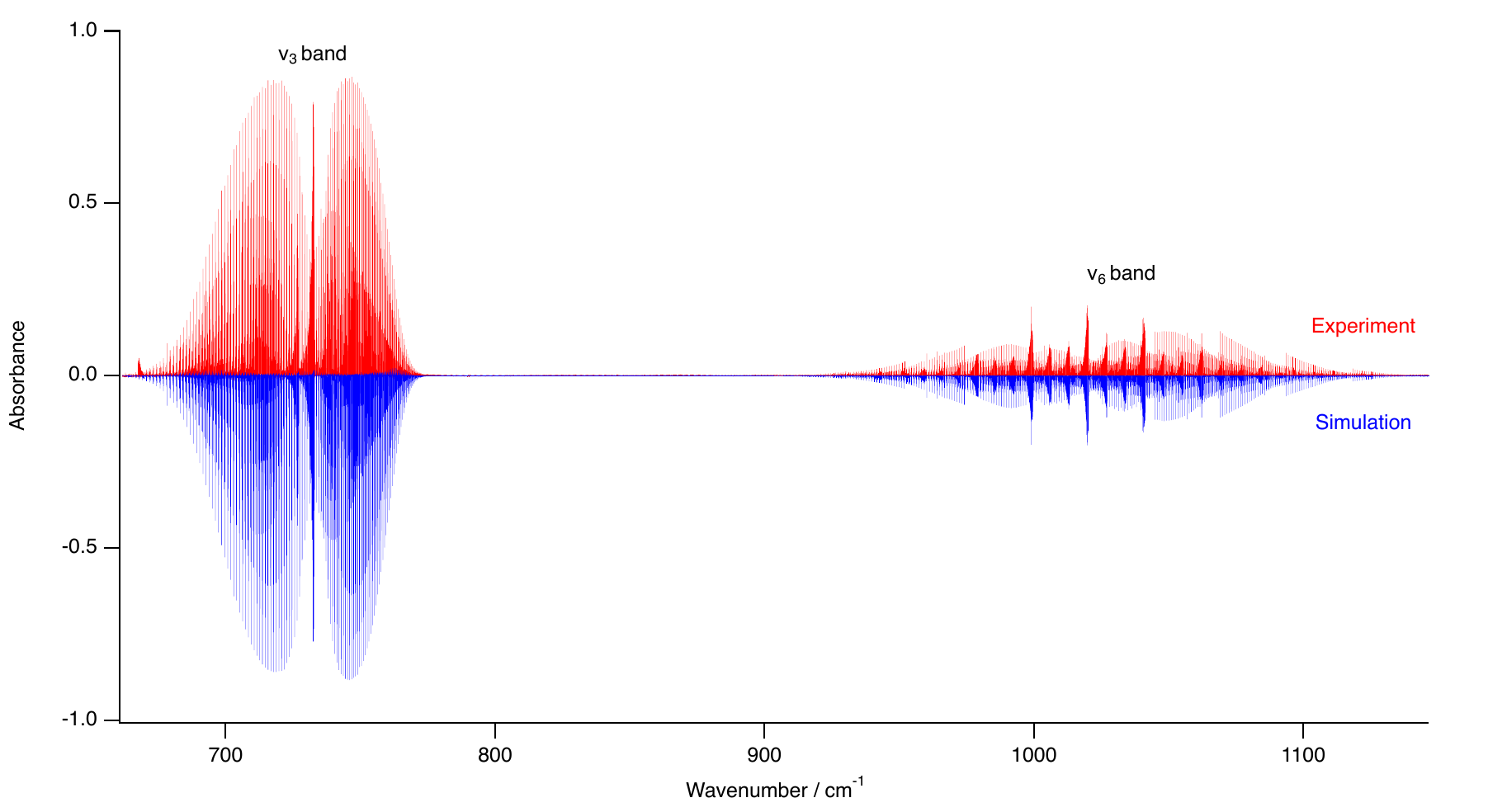}
    \caption{Comparison between experimental (red) and simulated (blue) spectrum of $\mathrm{CH_{3}Cl}$.}
    \label{fig:global}
\end{figure*}

\begin{figure*}[h!]
    \centering
    \includegraphics[width=\linewidth]{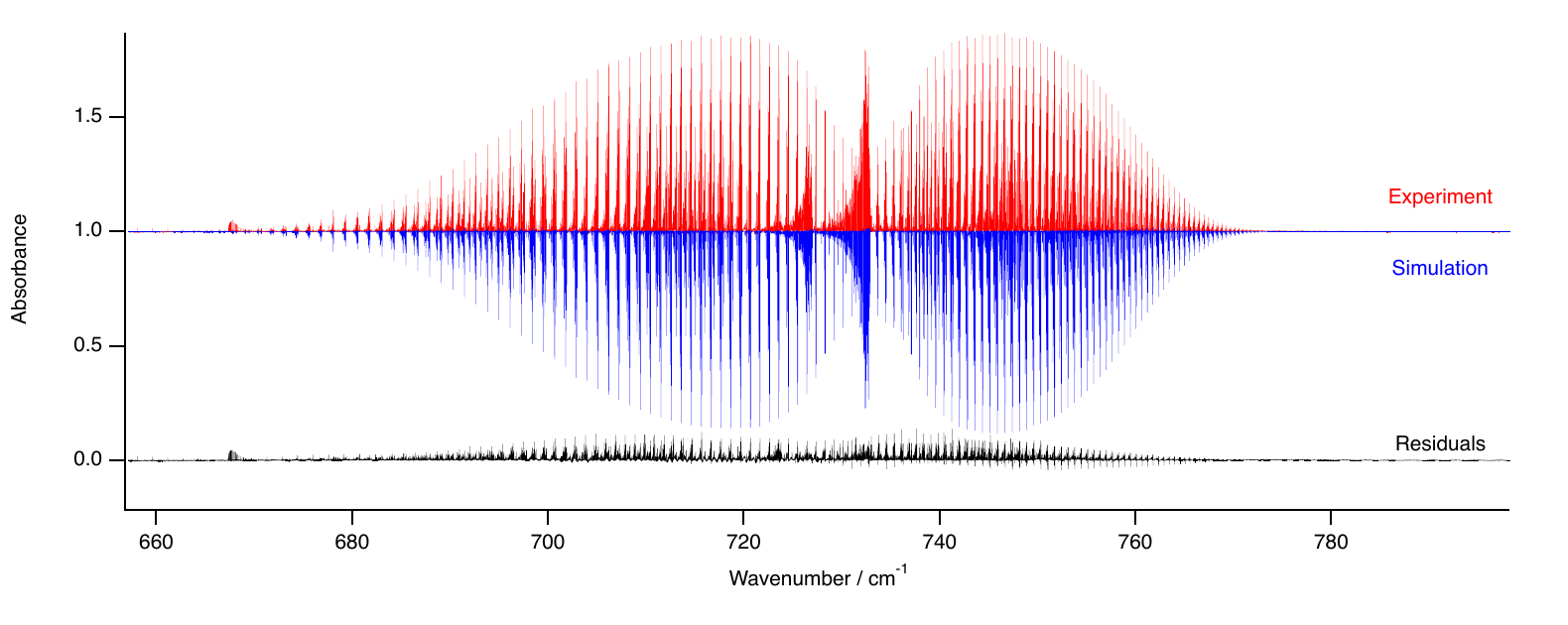}
    \caption{Comparison between experimental (red) and simulated (blue) $\nu_{3}$ band of $\mathrm{CH_{3}Cl}$. The difference is depicted in black.}
    \label{fig:nu3}
\end{figure*}

\begin{figure*}[p]
    \centering
    \includegraphics[width=\linewidth]{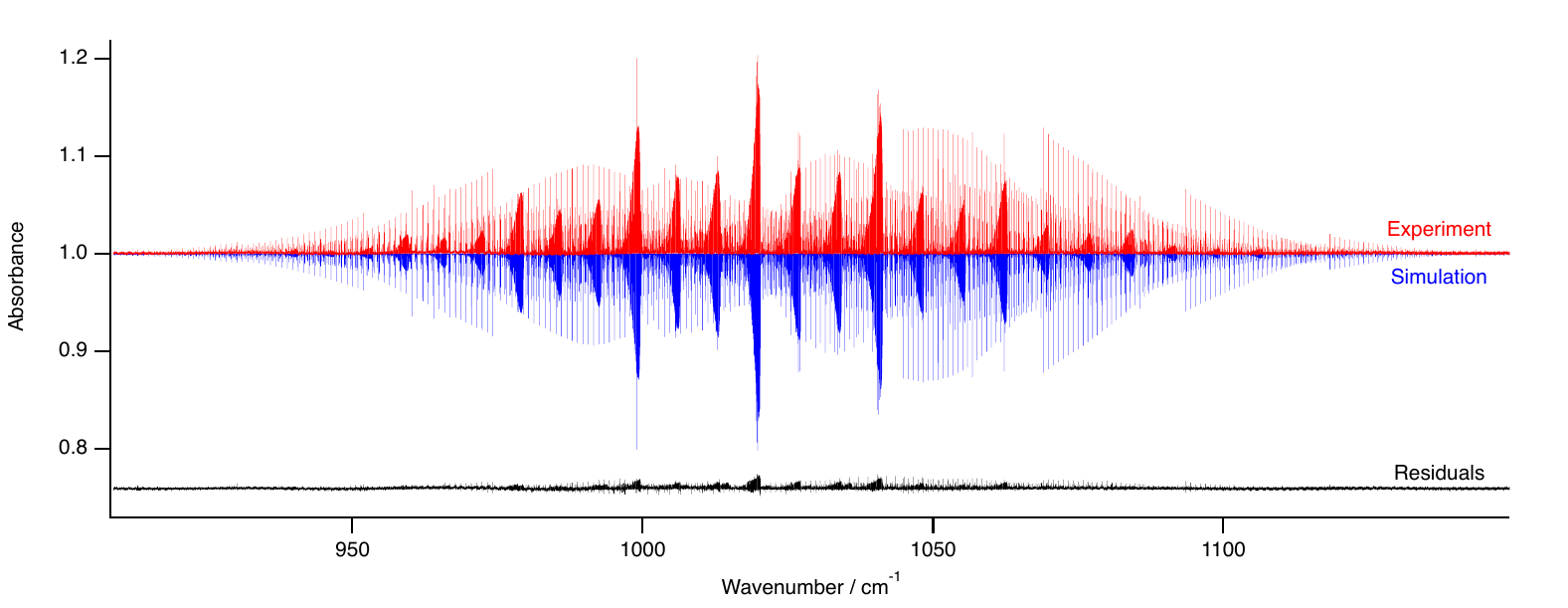}
    \caption{Comparison between experimental (red) and simulated (blue) $\nu_{6}$ band of $\mathrm{CH_{3}Cl}$. The difference is depicted in black.}
    \label{fig:nu6}
\end{figure*}

\begin{figure*}[p]
    \centering
    \includegraphics[width=\linewidth]{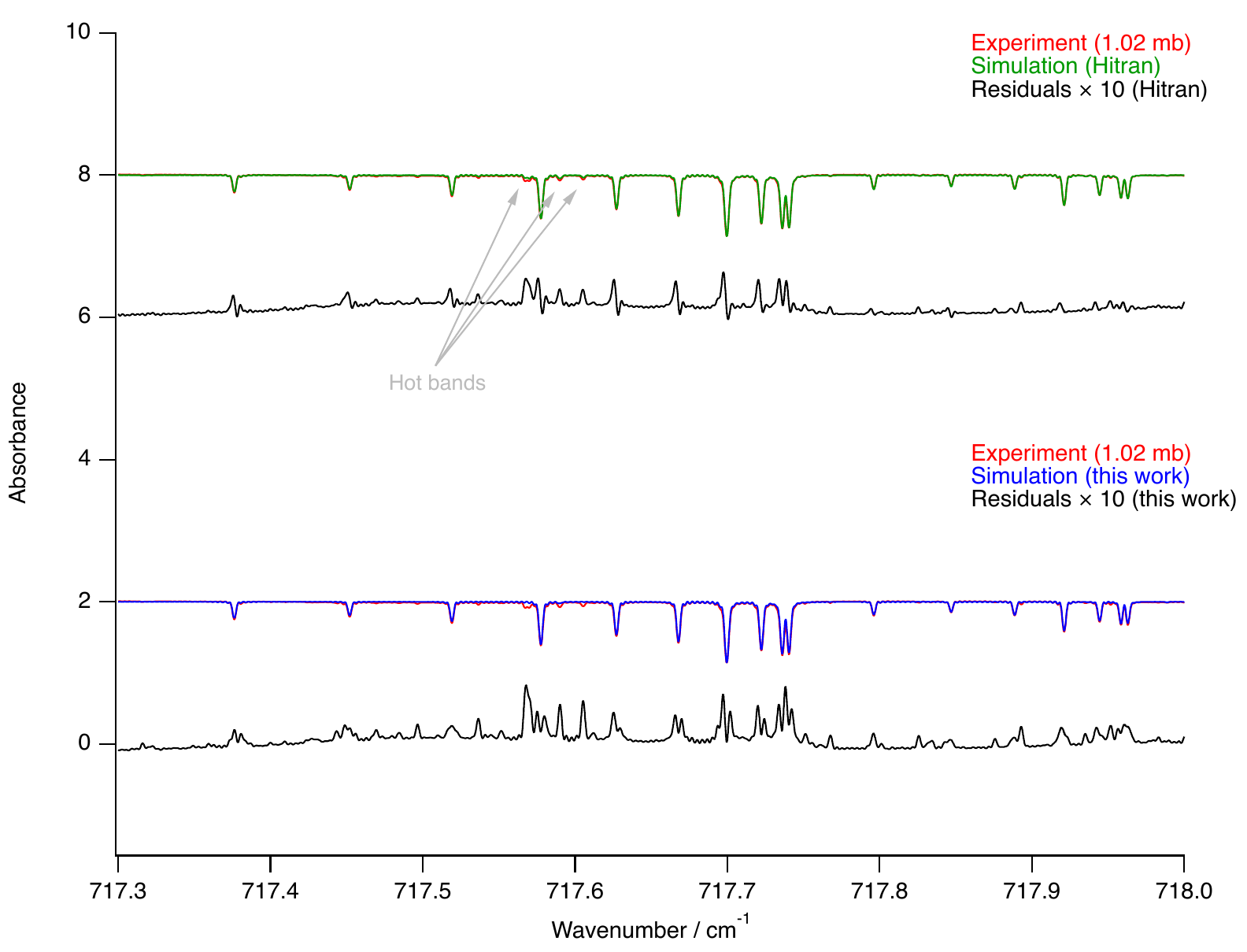}
    \caption{Zoom on a region of the $\nu_3$ band of $\mathrm{CH_{3}Cl}$. The spectrum in the upper part was simulated using the HITRAN database~\cite{hitran}, taking into account the hot bands. Our simulation is represented in the bottom part of the figure. Residuals are represented in black for both cases. They have been multiplied by a factor 10 to emphasize the presence of hot bands that were not simulated in our work. \color{black}Some hot bands lines are indicated by grey arrows in the HITRAN spectrum.}
    \label{fig:hotbands}
\end{figure*}

Spectra were simulated using the SGEN software developed by Jean Vander Auwera at Université Libre de Bruxelles, by taking into account the temperature (296~K) and pressure (1.02~hPa), and applying a Voigt line profile.
The experimental and simulated spectra are represented in Figures~\ref{fig:global}, \ref{fig:nu3}, and \ref{fig:nu6}. In Figure~\ref{fig:hotbands}, we compare the experimental spectrum with a simulation using the HITRAN database~\cite{hitran} and with our simulated spectrum.
Residuals are represented in both cases. The main difference between the two spectra is the presence of hot bands in the experimental spectrum that were not simulated in this work. Three of those bands can clearly be identified in the experimental and HITRAN spectra. Thanks to the HITRAN database \cite{hitran}, we determined that hot bands are mainly due to $2\nu_3-\nu_3$ transitions in this region.  We chose to restrict our analysis to the fundamental bands because our work is focused on the potential detection of this molecule in cometary atmospheres, which are very cold. Therefore we don't expect to observe those hot bands in cometary spectra.\color{black}

\newpage

\subsection{Derived Watson parameters}
Using the fitted tensorial parameters and equations of section~\ref{sec:formalism}, we derived the corresponding Watson parameters for both isotopologues using a conversion described in~\cite{willis}. We compared our derived parameters with the values presented in~\cite{DEMAISON1994147}. The Watson parameters are summarized in Table~\ref{tab:watpara}. Our obtained values are in very good agreement with the reference and more precise.
\begin{table*}[h]
\caption{Derived Watsonian parameters compared with values of Demaison {\em et al.\/}, 1994~\cite{DEMAISON1994147}. The values of C were not found in the reference, because the symmetry of the molecule prevents to derive it.}
\resizebox{\textwidth}{!}{
\centering
\begin{tabular}{lllll}
\hline \hline
\multicolumn{1}{l}{}                     & \multicolumn{2}{c}{$\mathrm{CH_3^{35}Cl}$}                           & \multicolumn{2}{c}{$\mathrm{CH_3^{37}Cl}$}                            \\ 
\multicolumn{1}{l}{Watsonian parameters} & \multicolumn{1}{l}{This work}             & Demaison {\em et al.\/}, 1994.~\cite{DEMAISON1994147}     & \multicolumn{1}{l}{This work}            & Demaison {\em et al.\/}, 1994.~\cite{DEMAISON1994147}       \\ \hline
C (MHz)                                    & \multicolumn{1}{l}{156063.259710(66)} &               & \multicolumn{1}{l}{156063.25954(54)} &                 \\
B (MHz)                                    & \multicolumn{1}{l}{13292.876670(66)}  & 13292.876660(60) & \multicolumn{1}{l}{13088.169470(540)}  & 13088.169630(380) \\
$D_j$ (kHz)                                   & \multicolumn{1}{l}{18.09611(42)}      & 18.09599(39)  & \multicolumn{1}{l}{17.56681(63)}     & 17.56713(41)    \\
$D_{jk}$ (kHz)                                  & \multicolumn{1}{l}{198.79898440(380)}   & 198.79900000(1480000) & \multicolumn{1}{l}{193.52197923(71)} & 193.52200000(2000000)     \\
$H_j$ (Hz)                                    & \multicolumn{1}{l}{-0.010130(640)}      & -0.010148(56) & \multicolumn{1}{l}{-0.009760(140)}     & -0.009732(59)   \\
$H_{jk}$ (Hz)                                   & \multicolumn{1}{l}{0.32870(460)}        & 0.32990(490)    & \multicolumn{1}{l}{0.31722(85)}      & 0.32510(530)      \\
$H_{kj}$ (Hz)                                   & \multicolumn{1}{l}{9.374400(5300)}        & 9.373000(147000)    & \multicolumn{1}{l}{9.284191(996)}    & 9.275000(169000)      \\ \hline \hline
\end{tabular}
}

\label{tab:watpara}
\end{table*}

\section{ChMeCaSDa database build}\label{sec:ChMeCaSDa}
We have set up a new database of calculated lines for the two isotopologues of CH$_3$Cl, ChMeCaSDa, in the framework of the VAMDC (“Virtual Atomic and Molecular Data Centre”)~\cite{dubernet2010virtual,dubernet2016virtual,moreau2018vamdc,atoms8040076}, on the same model as for our previous databases concerning CH$_3$D~\cite{RBR20}. The complete database is illustrated in Table~\ref{tab:database}, where $P_0$ is
the ground state, $P_1$ contains the $\nu_3$ band and $P_2$ the $\nu_6$. An amount of 12\,152 transitions have been included in the database.

The calculated data are accessible through our website at \url{http://vamdc.icb.cnrs.fr/PHP/CH3Cl.php}. Our webpage allows to plot the data and download two sorts of file formats: the line by line list is given following a hybrid of the HITRAN 2004 format, while cross section is a simple 2-column flat file. Indeed, we do not strictly follow the HITRAN 2004 format because our formalism does not involve the same quantum numbers. Therefore, the format used for the local quanta corresponds to the group 3 that belong to the spherical tops group (as explained in Ref.~\cite{rothman2005hitran}). The same choice was made for the methane database (MeCaSDa) with the CH$_3$D isotopologue.
\section{Comparison between ChMeCaSDA and HITRAN}\label{sec:Comparison}
\begin{figure}[p]
\centerline{\includegraphics[width=1.0\textwidth]{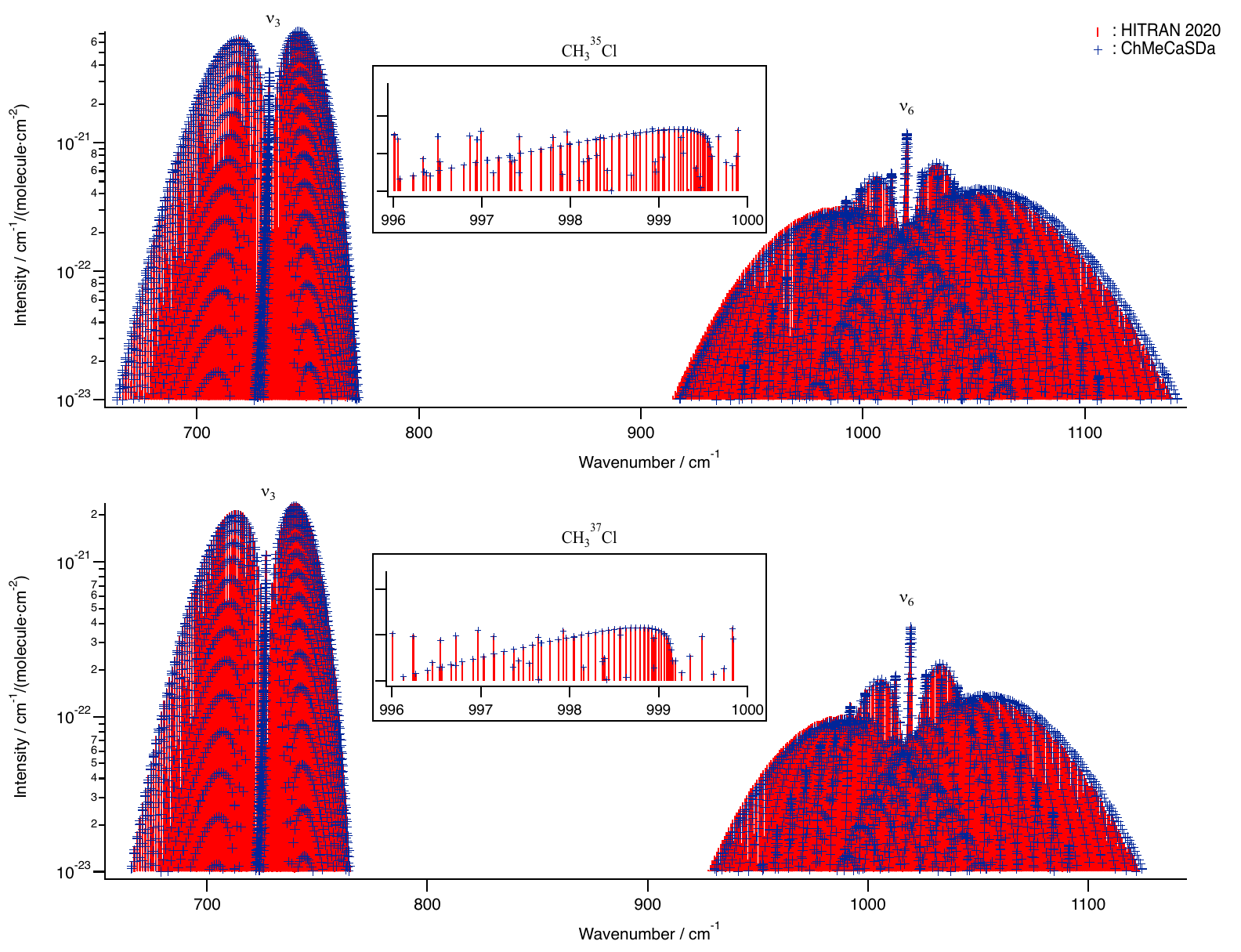}}
\caption{Comparison between the HITRAN (red) and ChMeCaSDa (blue, \url{https://vamdc.icb.cnrs.fr/PHP/CH3Cl.php}) databases for both isotopologues. The lines intensities of the ChMeCaSDa database were adjusted by taking into account the isotopic abundances of $\mathrm{CH_{3}Cl}$ (0.7577 for $\mathrm{CH_3^{35}Cl}$, 0.2423 for $\mathrm{CH_3^{37}Cl}$). Our database agrees very well with the data from HITRAN, for positions and intensities. Contrary to the HITRAN database, line intensities were fitted using the tensorial formalism in our new database. This proves once more the high reliability of this formalism, applied this time to the dipole moment. For both isotopologues, a small region of the $\nu_6$ band is included in the figure to better illustrate the agreement between the two databases.} 
\label{fig:ChMeCaSDa}
\end{figure}

An overview of the ChMeCaSDa database is displayed in blue in Fig.~\ref{fig:ChMeCaSDa}, as downloadable from our website. Data from the HITRAN database is represented on the same figure, in red.   In comparison with HITRAN, no new transitions were predicted. The number of lines is lower in ChMeCaSDa, because we chose a lower maximum rotational quantum number $J_{Max}$ in our simulation. This decision was once again taken in the context of the study of cold astrophysical environments. However, we significanlty improved the standard deviation of the lines positions ($3\times 10^{-4}\mathrm{cm^{-1}}$ in the HITRAN database of 2008 \cite{ROTHMAN2009533}, obtained from the work of Nikitin \textit{et al.} \cite{NIKITIN2005174}, using the same tensorial formalism). Concerning the intensities, they were also calculated using a tensorial formalism, as described in section \ref{sec:int}, whereas in HITRAN they were only adjusted thanks to experimental data, in the 2012 update~\cite{ROTHMAN20134}. 

\begin{table}[p]
\caption{Rovibrational transitions in ChMeCaSDa.}
\label{tab:database}
\begin{center}
\small\begin{tabular}{lrcc}
\hline\hline
Transitions     & Nb. dipolar  & Dipolar wavenumber    & Dipolar intensity                \\
                &              & cm$^{-1}$      & cm$^{-1}$/(molecule cm$^{-2}$)            \\
\hline
CH$_3^{35}$Cl   &              &                &                                            \\
$P_1-P_0$       &       2601   &  663 -- 773    & 8$\times10^{-24}$ -- 8$\times10^{-21}$    \\
$P_2-P_0$       &       3460   &  915 -- 1146   & 8$\times10^{-24}$ -- 2$\times10^{-21}$    \\
&               &              &                                            \\
CH$_3^{37}$Cl   &              &                &                                            \\
$P_1-P_0$       &       2593   &  658 --  767   & 8$\times10^{-24}$ -- 8$\times10^{-21}$    \\
$P_2-P_0$       &       3498   &  913 -- 1145   & 8$\times10^{-24}$ -- 2$\times10^{-21}$    \\
                &              &                &                                            \\
Total           &       12152  &                &                                            \\

\hline\hline
\end{tabular}
\end{center}
\end{table}

\color{black}

\clearpage

\section{Conclusion and perspectives}\label{sec:ccl}
Two strong fundamental absorption bands of chloromethane ($\nu_3$ and $\nu_6$) were recorded at high spectral resolution at the SOLEIL synchrotron facility. We successfully analyzed those bands using the Dijon tensorial formalism of symmetric molecules, for $\mathrm{CH_3^{35}Cl}$ and $\mathrm{CH_3^{37}Cl}$. We performed fits for line position and intensities. For the lines position, 23 tensorial parameters were derived for both isotopologues.

Our tensorial parameters converted in the Watson formalism are in very good agreement with the previous study~\cite{DEMAISON1994147}, providing more precise parameters than those found in the literature thanks to the use of larger datasets.

Spectra were simulated with the obtained fit parameters, taking into account the self-broade\-ning coefficients of both bands, previously reported in~\cite{DRIDI2019108,FATHALLAH2020106777}. With the exception of hot bands being only present in the experimental spectra, the simulated spectra are very close to the experimental ones.

Chloromethane is the first molecule analyzed in the frame of the COSMIC project (COmputation and Spectroscopy of Molecules in the Infrared for Comets), devoted to the study of molecules of cometary interests. Indeed, the only cometary detection of $\mathrm{CH_{3}Cl}$ was done \textit{in situ}, in the coma of 67P/Churyumov-Gerasimenko.

While the two bands analyzed in this work are not observable in the space from the ground because of the atmospheric absorption, we are planning to observe and analyze the $\nu_1$ band in the next few months. This band, absorbing around 2950~$\mathrm{cm^{-1}}$, occurs in the L infrared window. It is then of great interest to study, as it may be observable in infrared spectra of comets.

Many more molecules detected by the ROSINA mass spectrometer, are yet to be detected remotely in the comets. From remote detection, abundances can be derived, and constraints on cometary chemistry could be set. Through the COSMIC project, we hope to have a better understanding of comets, and therefore of the origins of the Solar system.

\section*{Acknowledgement}
The authors are grateful to Jean Vander Auwera from Université Libre de Bruxelles for providing the SGEN program to simulate the spectra presented above.
The COSMIC project is financed by the EIPHI Graduate School and held by ICB and UTINAM, supported by the {\em Conseil Régional de Bourgogne Franche-Comté\/} and the French National Research Agency (ANR). The authors wish to thank the anonymous referees for constructive feedback.


\clearpage
\bibliographystyle{unsrt}

\end{document}